\documentclass[10pt,aps,prb,twocolumn,showpacs,superscriptaddress]{revtex4-1}
\usepackage{amsmath,amssymb,setspace}
\usepackage{graphicx}

\begin{document}
\title{Magnetic order of intermetallic FeGa$_{3-y}$Ge$_y$ studied by $\mu$SR and $^{57}$Fe M\"ossbauer spectroscopy}

	\author{J. Munevar}
	\email[E-mail: ]{julian.munevar@psi.ch}
	\affiliation{Laboratory for Muon Spin Spectroscopy, Paul Scherrer Institut, 5232 Villigen PSI, Switzerland}
	\author{M. Cabrera-Baez}
	\affiliation{CCNH, Universidade Federal do ABC (UFABC), Santo Andr\'e, SP 09210-580, Brazil}
	\author{M. Alzamora}
	\affiliation{Universidade Federal do Rio de Janeiro, Campus Xer\'em, RJ 25245-390, Brazil}
	\author{J. Larrea}
	\author{E. M. Bittar}
	\author{E. Baggio-Saitovitch}
	\affiliation{Centro Brasileiro de Pesquisas F\'isicas, Rio de Janeiro, RJ 22290-180, Brazil}
	\author{F. J. Litterst}
	\affiliation{Centro Brasileiro de Pesquisas F\'isicas, Rio de Janeiro, RJ 22290-180, Brazil}
	\affiliation{Technische Universit\"at Braunschweig, 38106 Braunschweig, Germany}
	\author{R. A. Ribeiro}
	\author{M. A. Avila}
	\affiliation{CCNH, Universidade Federal do ABC, Santo Andr\'e, SP 09210-580, Brazil}
	\author{E. Morenzoni}
	\affiliation{Laboratory for Muon Spin Spectroscopy, Paul Scherrer Institut, 5232 Villigen PSI, Switzerland}

\date{\today}
	
\begin{abstract} Temperature dependent magnetization, muon spin rotation and $^{57}$Fe M\"ossbauer spectroscopy experiments performed on crystals of intermetallic FeGa$_{3-y}$Ge$_{y}$ ($y=0.11,0.14,0.17,0.22,0.27$, $0.29,0.32$) are reported.  Whereas at $y=0.11$ even a sensitive magnetic microprobe such as $\mu$SR does not detect magnetism, all other samples display weak ferromagnetism with a magnetic moment of up to 0.22 $\mu_B$ per Fe atom.  As a function of doping and of temperature a crossover from short range to long range magnetic order is observed, characterized by a broadly distributed spontaneous internal field.  However, the $y=0.14$ and $y=0.17$ remain in the short range ordered state down to the lowest investigated temperature.  The transition from short range to long range order appears to be accompanied by a change of the character of the spin fluctuations, which exhibit spin wave excitations signature in the LRO part of the phase diagram.  M\"ossbauer spectroscopy for $y=0.27$ and 0.32 indicates that the internal field lies in the plane perpendicular to the crystallographic $c$ axis.  The field distribution and its evolution with doping suggest that the details of the Fe magnetic moment formation and the consequent magnetic state are determined not only by the dopant concentration but also by the way the replacement of the Ga atoms surrounding the Fe is accomplished. \end{abstract}
\maketitle

\section{Introduction}

Tuning a material's physical properties by chemical doping or by application of some external control parameter such as high pressure or magnetic field often changes its behavior and induces new and exotic states of matter \cite{dagotto}.  A recent example is the electron doping of the intermetallic FeGa$_3$ that leads to enhanced thermoelectric figures of merit  \cite{hausermann,amagai,thermoelectric1,thermoelectric2,thermoelectric3,thermoelectric4,thermoelectric5,thermoelectric6,yin} and to emergent magnetic behavior accompanied by the possible observation of a Ferromagnetic Quantum Critical Point (FMQCP) \cite{umeo,haldolaarachchige,singh,michael,majumder,gamza,verchenko,gippius,alvarez,botana,likhanov}.  

FeGa$_3$ is a semiconductor with tetragonal structure (space group $P4_2/mnm$)\cite{hausermann} and a narrow band gap of approximately 0.5 eV caused by the hybridization of the 3$d$ Fe and 4$p$ Ga bands \cite{hausermann,thermoelectric2,thermoelectric6,bittar,osorio,gamza}.  It is diamagnetic over a broad temperature range and has a small Sommerfeld coefficient ($\gamma=$0.03 $\frac{\text{mJ}}{\text{mol K}}$) \cite{thermoelectric2,umeo,gamza}.  The Fe atoms occur in dimer pairs oriented along the $a$ and $b$ directions.  A unit cell contains 4 formula units where each Fe has eight Ga neighbors at two distinct sites Ga1 (0.236 nm, 2 atoms) and Ga2 (0.239 nm, 2 atoms and 0.246 nm, 4 atoms, above the plane containing Fe)\cite{hausermann}.  Whereas hole doping by Zn at the Ga site or Mn at the Fe site \cite{michael} does not induce an insulating-metal transition and introduces in-gap states \cite{gamza}, electron doping either at the Fe or the Ga site destroys the semiconducting behavior, and remarkably influences other physical properties \cite{umeo,haldolaarachchige,singh,michael,majumder,bittar,osorio,gamza,verchenko,gippius,alvarez,botana,likhanov,mondal}.  

Electron doping via Co substitution of Fe induces a shift of the Fermi level towards the conduction band, that leads to metallic-like transport and Curie-Weiss behavior already at low Co concentrations \cite{bittar}.  Large Co doping induces substantial disorder as reflected by the line broadening of the $^{69,71}$Ga Nuclear Quadrupole Resonance (NQR) spectra and by the deviation of the lattice parameters from Vegard's law \cite{verchenko}.  Fe$_{1-x}$Co$_x$Ga$_3$ remains paramagnetic for all Co concentrations investigated \cite{umeo}, while showing a complex magnetic behavior including itinerant and localized moment character and strong antiferromagnetic (AFM) spin fluctuations for Co substitution close to 0.5 \cite{gippius}.

In contrast, electron doping by substituting Ga with Ge in FeGa$_3$ has more dramatic effects on the magnetic properties.  It first suppresses the semiconducting and diamagnetic properties, and induces metallic and paramagnetic behavior at a Ge doping as low as $y=0.006$ \cite{umeo}.  Already at a low critical concentration $y_c=0.13-0.15$ a weak ferromagnetic (FM) state appears \cite{umeo,majumder} displaying features of non-Fermi liquid behavior \cite{haldolaarachchige}.  $^{71}$Ga NQR measurements, while not evidencing intrinsic structural disorder related to the Ge doping, point to an evolution from a correlated local moment metal at low Ge doping to a weakly itinerant 3D-ferromagnetism and indicate a crossover from short range to long range magnetic order\cite{majumder}.  For $y=0.15$ the divergence in $\frac{1}{T_1T}$ at $T=0$ K indicates very pronounced and pure 3D quantum critical fluctuations whereas the $y=0.2$ data can be well fitted within the self-consistent renormalization (SCR) theory \cite{majumder,moriya}.  The FM quantum critical behavior is manifested also by a temperature dependence of the specific heat and of M/H as that predicted by the SCR theory for FM spin fluctuations in three-dimensional systems \cite{umeo}.  

In spite of several investigations, experimental as well as theoretical, the nature and evolution of the magnetic order in FeGa$_{3-y}$Ge$_y$ is far from being well understood.  Magnetism in FeGa$_{3-y}$Ge$_y$ has been discussed in terms of itinerant magnetism, of local magnetic moments or of a combination of both.  The itinerant view is supported by the small saturated moment and corresponding large Rhodes-Wohlfarth ratio \cite{umeo,haldolaarachchige}.  DFT calculations in a weakly correlated picture find that itinerant magnetism in FeGa$_3$ can be obtained by modest electron (but also hole) doping, without the presence of preformed moments.  The density of states increasing very rapidly with narrow bands near the band edges  suggests the possibility of a Stoner mechanism of ferromagnetism when doped \cite{singh}.  Botana \emph{et. al.} compared results from weakly and strongly correlated pictures and found that in both cases magnetism including itinerant phases appears easily with doping \cite{botana}.  

Recent DFT calculations supported by some magnetic susceptilibity measurements have suggested a complex development of the magnetism of Ge doped FeGa$_3$ with a gradual evolution from localized moments to a more delocalized character state and a combination of localized and itinerant moments accompanied by interplay of ferromagnetism and antiferromagnetism until itinerant magnetism is established at high doping level of about $y=0.4$.  This behavior appears to depend not only on the dopant concentration but also on the local Ge configuration with respect to Fe \cite{alvarez}.


The rich and complex behavior including magnetism and quantum critical behavior observed by partial substitution in FeGa$_3$ calls for investigations that are sensitive to the local Fe environment.  We present here muon spin rotation/relaxation ($\mu$SR) and $^{57}$Fe M\"ossbauer spectroscopy measurements on FeGa$_{3-y}$Ge$_y$ as function of Ge doping.  These techniques are able to give unique information about the local magnetic fields, field distribution and fluctuations at the muon and the Fe site, respectively.  In particular $\mu$SR is sensitive to static and dynamic spin correlations in systems with critical behavior \cite{adroja,spehling,amato} and can determine the degree of homogeneity of the magnetic phase and how it develops with temperature.  $^{57}$Fe M\"ossbauer spectroscopy gives information about the electric and magnetic surrounding of Fe at a lattice position.

Our results indicate an evolution from short range to long range order magnetism, displaying near $y_c$ a large degree of inhomogeneity with peculiarities that can be related to the magnetic moment distribution of the Fe atoms.  The spin fluctuations appear to have different character close to the QCP when compared to that of the well established FM phase.

\section{Experimental Details}

Single crystalline specimens of FeGa$_{3-y}$Ge$_y$ ($y=0.11,0.14,0.17, 0.22, 0.27, 0.29, 0.32$) have been grown at UFABC using the Ga self flux route \cite{canfield,ribeiro}.  High purity elements were sealed in an evacuated quartz ampoule and heated in a box furnance to 1100 $^\circ$C and then slowly cooled to 550 $^\circ$C over 150 h. More details on single crystal growth of these materials are provided elsewhere \cite{michael}.  The effective Ge concentration $y$ of the samples were estimated using energy dispersive X-ray spectroscopy (EDS) measurements in a JEOL model JSM-6010LA scanning electron microscope with a Vantage EDS system.  The estimated $y$ gives effective moment and transition temperature compatible with reported results\cite{umeo}.

Magnetization measurements in the 2 - 300 K temperature range were performed for each sample in a MPMS Quantum Design Magnetometer (SQUID-VSM) under an applied field H = 1 T (Fig. \ref{magnetization}).  For all Ge concentrations, zero field (ZF) and weak transverse field (wTF) muon spin rotation spectra were obtained at the GPS and Dolly, at the Swiss Muon Source of the Paul Scherrer Institut, Switzerland.  For $y=0.14$ we performed selected pressure dependent $\mu$SR measurements at the GPD instrument.  Temperature dependent $^{57}$Fe M\"ossbauer spectra (MS) for  FeGa$_{2.73}$Ge$_{0.27}$ and FeGa$_{2.68}$Ge$_{0.32}$ were obtained at the Brazilian Center for Research in Physics (CBPF), Brazil, by recording the energy dependent $\gamma$-ray transmission on powdered specimens of the above mentioned single crystals.  A 14.4 keV $\gamma$-ray radiation source of $^{57}$Co in Rh matrix delivering about 50 mCi, kept at the same temperature of the absorber, and a standard transmission spectrometer with sinusoidal velocity sweep were used.  The temperature ranges for the $\mu$SR and MS measurements were from 0.25 K to 300 K and from 4.2 K to 300 K, respectively.

\section{Results}

\subsection{Magnetization Results}

\begin{figure}[h!]
 \centering
 \includegraphics[width=9cm]{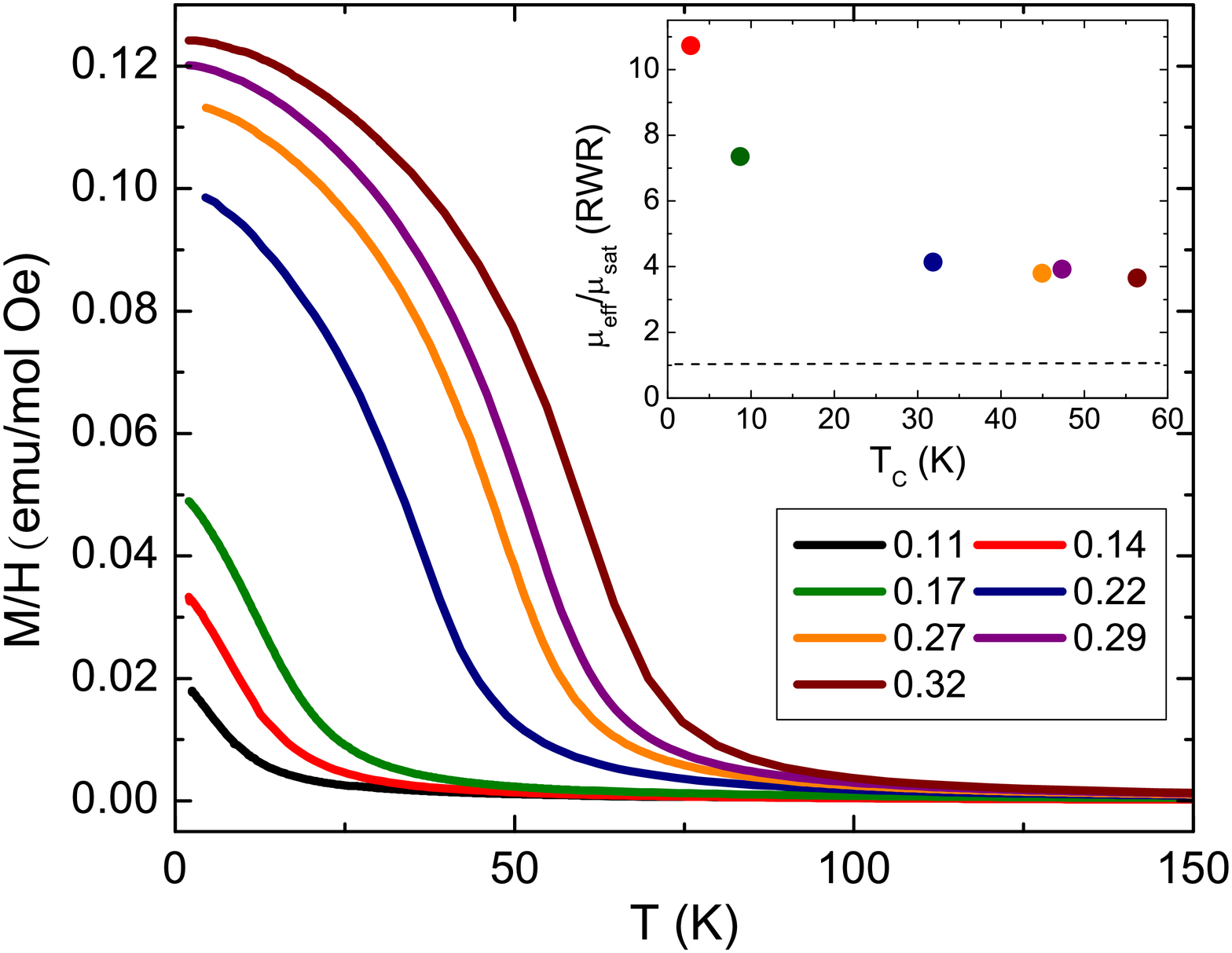}
 \caption{Temperature dependent ZFC magnetization measurements for FeGa$_{3-y}$Ge$_y$. single crystals in a 1 T external field.  In the inset the Rhodes-Wohlfarth ratios obtained for each sample are shown.}\label{magnetization}
\end{figure}

The magnetic response M/H as a function of temperature (Fig. \ref{magnetization}) clearly reflects the FM nature of FeGa$_{3-y}$Ge$_y$.  The ferromagnetic moment ranges from 0.09 $\mu_B$ per Fe atom for the $y=0.17$ sample, to 0.22 $\mu_B$ per Fe atom for $y=0.32$, significantly smaller than that of pure Fe (2.22 $\mu_B$).  The inset of Fig. \ref{magnetization} shows the Rhodes-Wohlfarth ratio (RWR=$\dfrac{\mu_{eff}}{\mu_{sat}}$) for different Ge concentrations, clearly above the expected value (RWR=1) for localized ferromagnetism \cite{umeo,haldolaarachchige,alvarez,zhang}.  

\begin{figure*}[!ht]
 \centering
 \includegraphics[width=18cm]{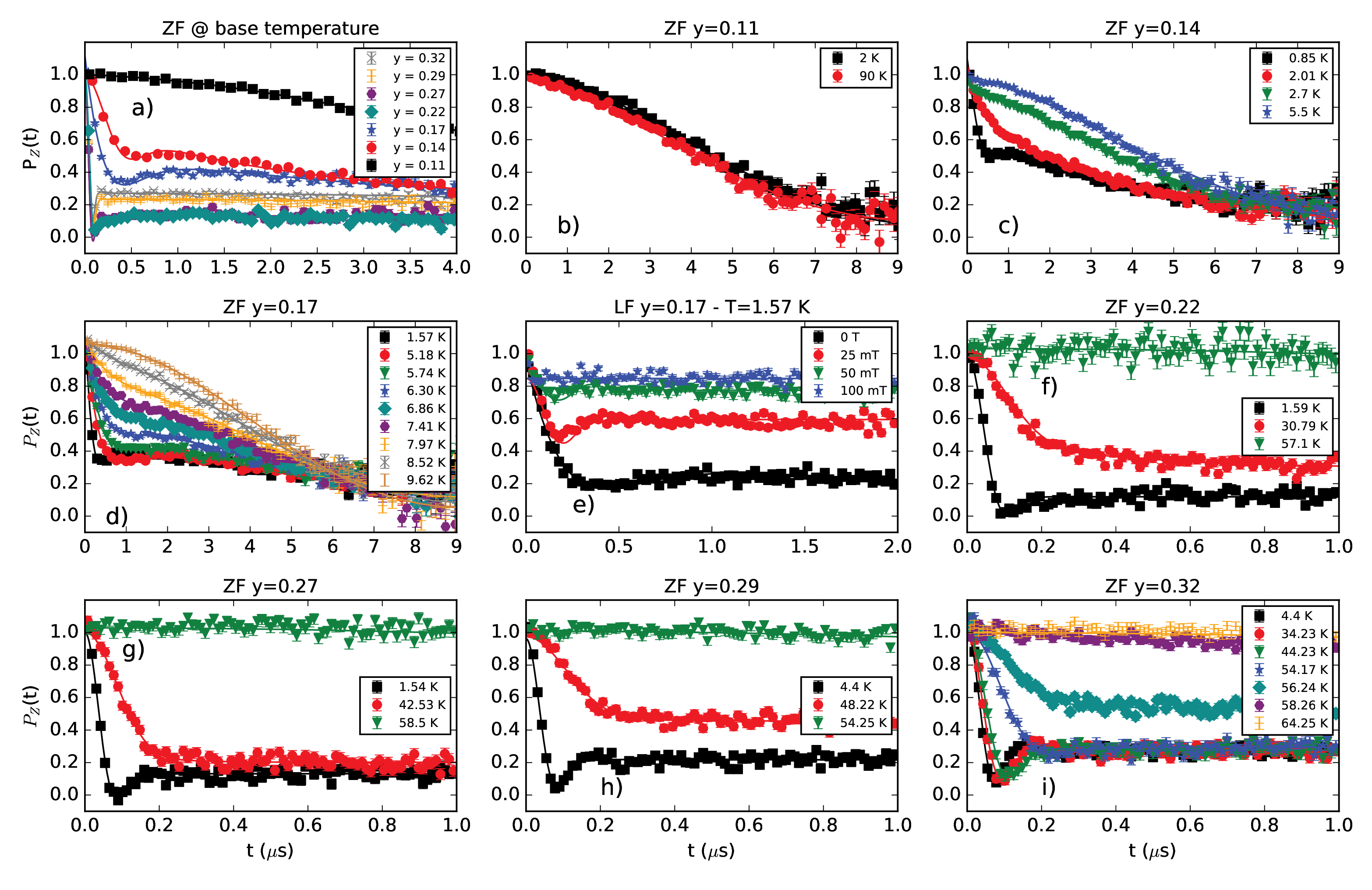}
 \caption{(a) Lowest temperature measurements of the muon spin polarization in ZF for all investigated Ge concentrations $y$. (b)-(d),(f)-(i) Temperature evolution of the ZF $\mu$SR spectra for FeGa$_{3−y}$Ge$_y$ at various $y$. (e) shows the LF spectra for $y$ = 0.17 at fields up to 100 mT. }\label{spectra}
\end{figure*}

\begin{figure*}[ht!]
 \centering
 \includegraphics[width=18cm]{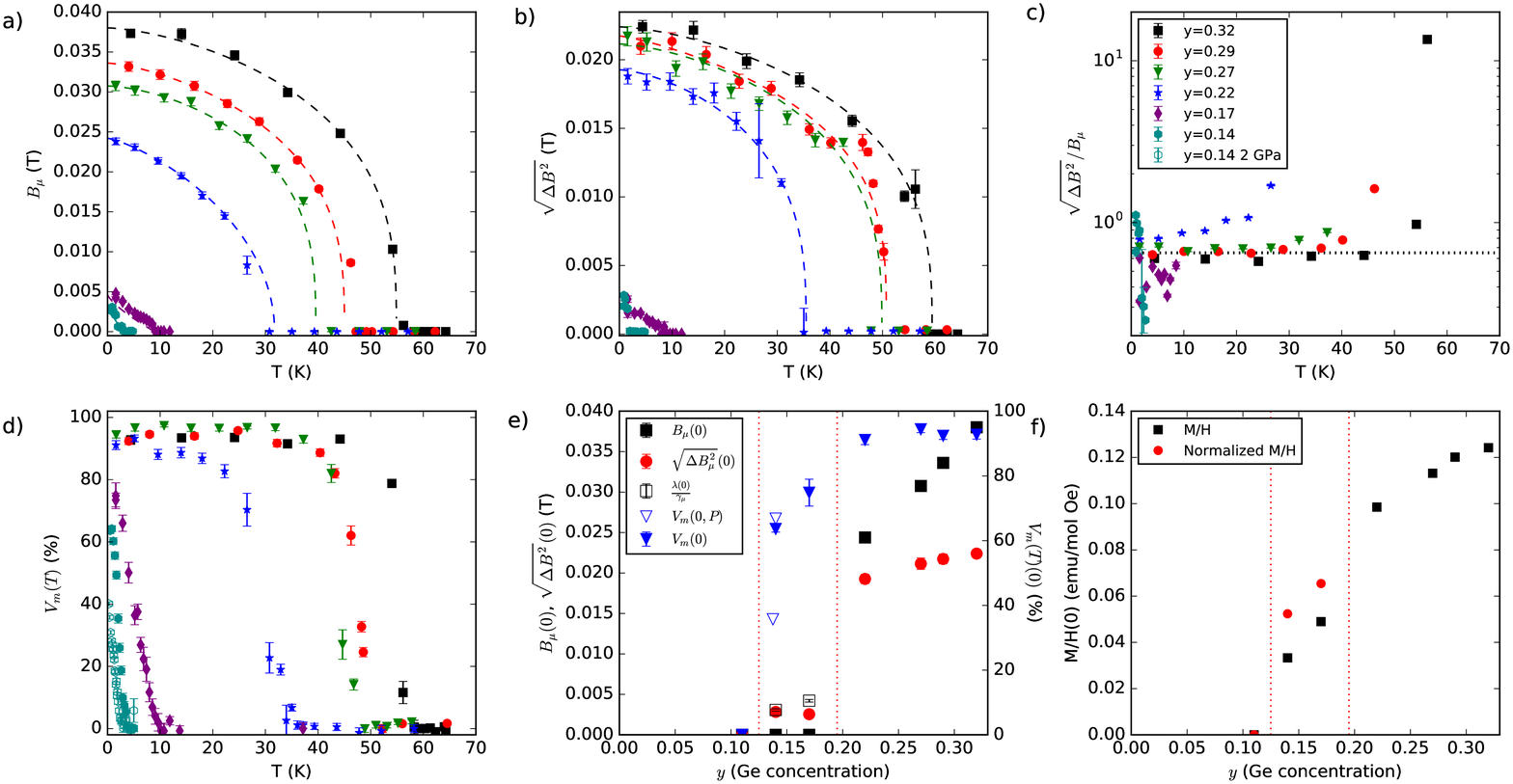}
 \caption{Doping dependence of various parameters obtained from the $\mu$SR measurements of FeGa$_{3-y}$Ge$_y$ single crystals.  Temperature dependence of (a) spontaneous internal field $B_{\mu}(T)$.  For $y=0.14$ and 0.17, where $B_{\mu}=0$, $\frac{\lambda}{\gamma_{\mu}}$ is plotted.  (b) Field width $\sqrt{\Delta B^2}(T)$, (c) ratio $\frac{\sqrt{\Delta B^2}}{B_{\mu}}$, (d) magnetic volume fraction as obtained from the analysis of the weak transverse field $\mu$SR spectra.  The corresponding points for the magnetic volume fraction under pressure is represented by hollow circles.  (e) Doping dependence of the low temperature limit of the spontaneous internal field $B_{\mu}(0)$, of the field width $\sqrt{\Delta B^2}(0)$ and of the magnetic volume fraction $V_m(0)$.  (f) Doping dependent saturation magnetization $M(0)/H$ obtained from Fig. \ref{magnetization}, and normalized to the magnetic volume fraction.}\label{fig2}
\end{figure*}

\subsection{$\mu$SR Results}

Selected ZF-$\mu$SR spectra from FeGa$_{3-y}$Ge$_y$ single crystals are shown in Fig. \ref{spectra}.  The evolution of the magnetism as a function of doping and across the critical concentration $y_c$ is already evidenced in Fig. \ref{spectra}(a).  Whereas the $y=0.11$ sample shows only weak temperature independent muon spin depolarization (see also Fig. \ref{fig2}(b)) with characteristic Gaussian Kubo-Toyabe behavior due to the static nuclear magnetic moments of Ga and Fe, the fast relaxation which sets in at early times on lowering the temperature for $y\geq 0.14$ reflects the onset of electronic magnetism.  The flat behavior of the polarization at late times for $y\gtrsim0.2$ reflects static magnetism.  The weak decay of this tail for $0.14\leq y \leq 0.17$ (see Fig. \ref{spectra}(c)-(d)) indicates the persistence of some slow ($\sim$MHz) spin fluctuations in the low doping range.  The (quasi) static nature of the magnetism is confirmed by longitudinal field data for $y=0.17$ (Fig. \ref{spectra}(e)), which shows the complete decoupling of the muon spin from the local field in a longitudinal field of 0.1 T. 

At high temperatures all the spectra display the typical muon spin relaxation behavior due to the very small static nuclear magnetic moments of Ga and Fe.  The muon polarization in this non-magnetic regime can be modeled as

\begin{equation}
 P_Z^{pm}(t)=\frac{1}{3}+\frac{2}{3}\left(1-\sigma_{n}^2t^2\right)\exp{\left(-\frac{\sigma_{n}t^2}{2}\right)},\label{eq1}
\end{equation}

where $\sigma_n$ is the Gaussian muon spin relaxation rate caused by nuclear moments.  The model that best describes the spectra at $y=0.14$ and $y=0.17$ is the sum of two sample volume contributions: a paramagnetic fraction $1-f$ described by Eq.\ref{eq1} and a magnetic one $f$ described by the so-called combined Kubo-Toyabe function:

\begin{align}
P_Z(t)&=(1-f)P_Z^{pm}(t)+f\left[\frac{1}{3}\exp{\left(-\lambda_{l}t\right)}+\right.&\nonumber\\&\left.+\frac{2}{3}\left(1-\lambda t-\sigma_T^2t^2\right)\exp{\left(-\frac{\sigma_T^2t^2}{2}\right)}\exp{\left(-\lambda t\right)}\right],\label{eq2}
\end{align}

where $\lambda$ is the Lorentzian muon spin relaxation rate, $\sigma_T=\gamma_{\mu}\sqrt{\Delta B^2}$ is the Gaussian muon relaxation rate, and $\lambda_l$ accounts for the small damping of the tail of the polarization, corresponding to slow dynamic fluctuations.  This accounts for the strong muon spin depolarization observed in Fig. \ref{spectra}, as being caused by two different sources of magnetism: a dense distribution of magnetic moments producing a Gaussian field distribution with variance $\frac{\sigma_T^2}{\gamma_{\mu}^2}$ ($\gamma_{\mu}=2\pi\times135.5$ MHz/T) in an environment of diluted magnetic moments producing a Lorentzian distribution with HWHM $\frac{\lambda}{\gamma_{\mu}}$.  Both distributions probed by the muons are centered around a local field $B_{\mu}$ with zero $x$, $y$, and $z$ components.  The value $\frac{1}{3}$ of the tail reflects the isotropic distribution of the local fields.

For larger doping $y>0.17$ the magnetic contribution to the data cannot be simply described by an isotropic distribution around $B_{\mu}=0$, instead a spontaneous field $B_{\mu}\neq 0$ at the muon site has to be taken into account.  The case of an isotropic Gaussian distribution around an isotropic static field of constant magnitude is known as the Koptev-Tarasov model \cite{musrbook,schroeder}, from which for $B_{\mu}=0$ the Kubo-Toyabe formula is easily recovered.  For not too small $B_{\mu}$ the polarization function can be described by Eq.\ref{eq4}, which is used to fit the low temperature data:

\begin{align}
 P_Z^m(t)&=a\exp{\left(-\lambda_l t\right)}+\nonumber\\&+(1-a)\exp{\left(-\frac{\sigma_T^2t^2}{2}\right)}\cos{\left(\gamma_{\mu} B_{\mu} t\right)},\label{eq4}
\end{align}

together with a temperature dependent contribution describing the paramagnetic fraction given by Eq. \ref{eq1}.  We have therefore for $P_Z(t)$

\begin{equation}
 P_Z(t)=(1-f)P_Z^{pm}+fP_Z^m(t),\label{eq5}
\end{equation}

where $P_Z^{pm}(t)$ and $P_Z^m(t)$ are the polarization functions in the paramagnetic and the magnetic state.  $B_{\mu}$ is the internal field sensed at the muon site, and the parameter $a$ is related to the fraction of muons with initial spin parallel to an internal field component.  Note that $a=\frac{1}{3}$ corresponds to an isotropic distribution of fields.  Since our sample consists of a set of single crystals, a deviation from $a=\frac{1}{3}$ indicates a preferred orientation of the local field with respect to the crystal axes. 

The temperature dependence of $B_{\mu}$ and of $\sqrt{\Delta B^2}$ are plotted in Fig. \ref{fig2}(a) and (b).  For the $y=0.14$ and $y=0.17$ samples, where $B_{\mu}=0$, we have plotted the width of the Lorentzian distribution $\frac{\lambda}{\gamma_{\mu}}$ as a measure of the local field strength.  Both $B_{\mu}(T\rightarrow 0)$ and $\sqrt{\Delta B^2}(T\rightarrow 0)$ as well as $T_{\rm C}$ increase as the Ge concentration increases, in agreement with the observed increase of the spontaenous magnetic moment saturation per formula unit\cite{umeo}.  However, the doping dependent low temperature values of the internal field and field width reflect a change of the magnetic regime at $y\sim0.2$ with a step like increase of both parameters.  For $y>0.2$ the temperature dependence of $B_{\mu}$ and $\sqrt{\Delta B^2}$ are similar and represent the build-up of the local order parameter of a magnetic transition of second order.

Interestingly, the ratio $\frac{\sqrt{\Delta B^2}}{B_{\mu}}$ (Fig. \ref{fig2}(c)), which is effectively infinite for $y=0.14$ and $0.17$ where $B_{\mu}=0$ and where also $\frac{\gamma_{\mu}\sqrt{\Delta B^2}}{\lambda}$ is large, remain quite large for $y>0.17$ reflecting a broad field distribution at all investigated Ge concentrations.  The relatively low values of $B_{\mu}(0)$ and its evolution with Ge doping are consistent with a weak FM state evolving from short range order for $y\lesssim0.17$ to more long range order with increasing concentration.  The parameter $a$ of Eq.\ref{eq4} ranges from 0.12 to 0.27 for $0.22\leq y \leq 0.32$.  The deviation from the value $\frac{1}{3}$ for the isotropic case indicates that the field has a preferred orientation with respect to the crystal axis, as also suggested by the M\"ossbauer spectra discussed in the following section.

We also determined the magnetic volume fraction $V_m(T)$ as a function of temperature from the precessing asymmetry measured in weak transverse field experiments of 5 mT.  The results are shown in Fig. \ref{fig2}(d).  For Ge doping $y\gtrsim 0.22$ the magnetic volume fraction is nearly 100 \%, shows a sharp transition at $T_{\rm C}$ at the highest doping and a small rounding below.  This together with the gradual increase of the local order parameter mentioned above is an additional signature of second order transition.  By contrast, at $y=0.14$ and 0.17, which are just above the critical Ge concentration, magnetic order develops gradually with temperature and reaches only about 70 \% of the sample volume even at the lowest temperature.  The magnetic volume fraction of $y=0.14$ under external pressure of 2 GPa shows a further decrease down to 40 \%, indicating the suppression of the magnetic ground state by pressure, which has been previously shown to induce a decrease in $T_{\rm C}$\cite{umeo}.

By plotting $B_{\mu}(0)$, $\sqrt{\Delta B^2}(0)$ and $V_m(0)$ versus Ge concentration, the effect of doping in the magnetic properties (Fig. \ref{fig2}(e)) is clearly seen.  A continuous decrease of $B_{\mu}(0)$ and $\sqrt{\Delta B^2}(0)$ down to Ge $y=0.22$ is observed, and $V_m(0)$ remains nearly constant in this range.  Lower Ge concentrations induce a dramatic decrease in $B_{\mu}(0)$, $\sqrt{\Delta B^2}(0)$, and $V_m(0)$ starts to decrease, and finally at $y=0.11$ no magnetic order is detected.  Fig. \ref{fig2}(f) shows the saturation magnetization $M(0)/H$ for each Ge concentration, and the corresponding quantity normalized $M(0)/H$ by $V_m(0)$ (Fig. \ref{fig2}(e)), which indicates a smoother development of the average effective magnetic moment with doping.

\subsection{$^{57}$Fe M\"ossbauer Spectroscopy Results}

\begin{figure}[h!]
 \centering
 \includegraphics[width=9cm]{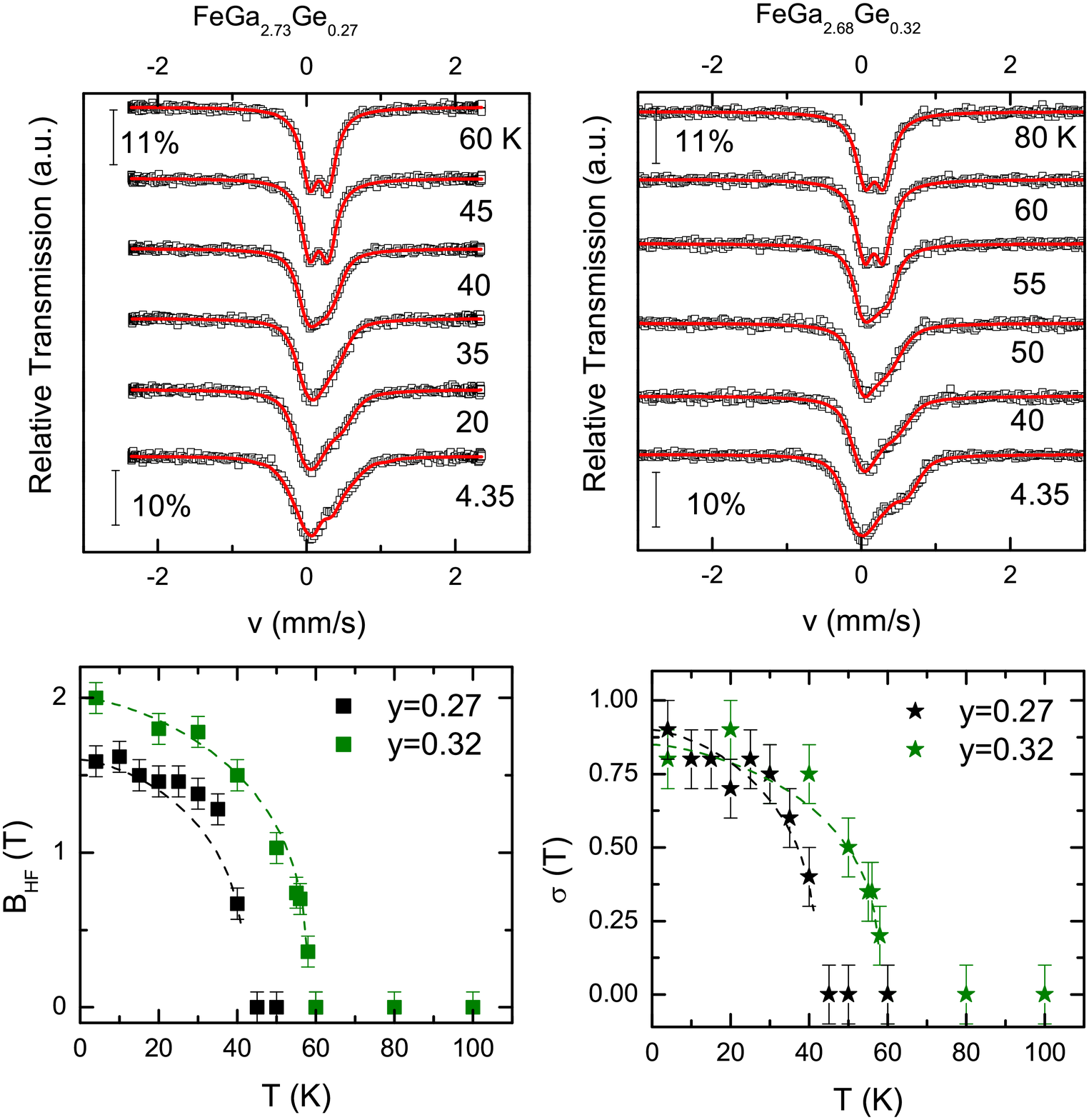}
 \caption{$^{57}$Fe M\"ossbauer spectra for $y=$0.27 and 0.32 crystals at different temperatures.  In the bottom part of the figure the temperature dpendence of the Fe hyperfine field and its distribution width are shown.}\label{figmoss}
\end{figure}

$^{57}$Fe M\"ossbauer absorption spectra on samples of FeGa$_{3-y}$Ge$_y$ ($y=$0.27 and 0.32) are shown in Fig. \ref{figmoss}.   A clear doublet profile is observed in the paramagnetic state, which was fitted with a nuclear electric quadrupole interaction between the iron nucleus and its surroundings.  When the temperature is lowered and the magnetically ordered regime is entered, the resonance lines broaden and the absorption profiles become asymmetric reflecting the effect of the magnetic hyperfine field $B_{hf}$ (see Fig. \ref{figmoss}).  These spectra have low resolution due to the small values of $B_{hf}$, therefore their analysis depends to some degree on the chosen procedure.  Obviously the strengths of nuclear electric quadrupole interaction and magnetic hyperfine interaction are comparable necessitating the solution of a full Hamiltonian comprising both interactions for the determination of line positions and their proper intensities.  We have employed the codes of MOSSWIN \cite{mosswin} and private ones \cite{litterst} allowing also for transmission integral corrections of spectral line shape.  

For the spectra of the absorber with $y=$0.27 taken in applied magnetic fields at 4.2 K (not shown) we had to assume a random orientation of an axial electric field tensor versus the direction of applied field.  For the sample with $y=$0.32 we could achieve an alignment of crystallites along the applied field as could be traced from the missing nuclear transitions with $\Delta I_z=$0 ($I_z$ being the nuclear spin projection) connecting the nuclear excited and ground states with spin $I_z=\frac{3}{2}$ and $I_z=\frac{1}{2}$, respectively.

Above the magnetic ordering temperatures the values of quadrupole splittings $eV_{ZZ}Q/2$ ($e$ is the elementary charge, $V_{ZZ}$ is the electric field gradient major component, $Q$ the nuclear quadrupole moment for $^{57}$Fe in its excited state) are practically equal with 0.25(3) mm/s for both $y=0.27$ and $0.32$.  Also, isomer shifts are equal, with $\delta_{IS}=$0.28(1) mm/s (vs. Fe metal at room temperature).  When entering the magnetically ordered state the derived values for the quadrupole interactions are only about half of those above $T_{\rm C}$, if it is assumed that the electric field gradient main component $V_{ZZ}$ is oriented along $B_{hf}$.  From the spectra in applied field on the oriented sample $y=$0.32, however, it becomes clear that the latter assumption is erroneous and instead the axial field gradient tensor is oriented perpendicular to $B_{hf}$ with a negative value $V_{ZZ}$.  This means that the nuclear electric quadrupole interaction in the paramagnetic and the ferromagnetic state is the same.  Assuming the main axis of the electric field gradient to be tetragonal $c$ axis, we have to conclude that $B_{hf}$ lies within the $ab$ plane.  This is in agreement with the $\mu$SR results which point to a preferred orientation of the internal fields.  For further analysis of the M\"ossbauer spectra in the magnetically ordered regime this was taken into account for both samples.  Whereas line widths in the paramagnetic regime do not reveal a noticeable broadening caused by a distribution of isomer shift and quadrupole interactions, we have to introduce a wide distribution of magnetic hyperfine splittings in the magnetic state.  Best fits were achieved with a Gaussian distribution width $\sigma$ around a mean value $B_{hf}$.

Mean magnetic hyperfine fields $B_{hf}$ and Gaussian widths $\sigma$ obtained from the fits described above are shown in Fig. \ref{figmoss}, following a similar behavior as the $\mu$SR internal fields and field widths $\sqrt{\Delta B^2}$ in Fig \ref{fig2}.  The saturation values of $B_{hf}$ measured at lowest temperatures are consistent with those derived from spectra obtained in applied magnetic fields.  

\section{Discussion}

The results obtained in the present investigation by local probe techniques show the development of weak FM upon electron doping of FeGa$_{3-y}$Ge$_y$, with an evolution of the character of the magnetic order on increasing $y$.  The presence of Fe atoms and their dimer arrangement with relative distance of 0.277 nm (to be compared with a nearest Fe-Fe distance of 0.248 nm in the bcc iron metal) has raised the question about the pre-existence of magnetic moments and their magnetic behavior even in the undoped compound FeGa$_3$, which shows diamagnetic properties. LDA calculations including a realistic on site repulsion have suggested an antiferromagnetic arrangement of Fe equivalent to a Fe$_2$ spin singlet state with a Fe magnetic moment of 0.6 $\mu_B$ and suggested that the doped induced magnetism would be linked to the breaking of the singlets into free spins \cite{yin}.   On the other hand, calculations by Singh within the GGA approximation explained the magnetism of doped FeGa$_3$ without resorting to the coupling of pre-existing spins. \cite{singh}

At the lowest doping investigated in this work ($y=0.11$), below the concentration where a FMQCP is expected, a very sensitive local magnetic probe such as $\mu$SR does not find indication of a magnetic state. The weak exponential relaxation, which appears on increasing the temperature in the ZF $\mu$SR spectra (see Fig. \ref{spectra}(b) may indicate a paramagnetic contribution related to the presence of some free magnetic moments. This appears difficult to reconcile with the antiferromagnetic order as the one lowest in energy calculated by Yin and Pickett \cite{yin} and with results of recent neutron powder diffraction measurements that found magnetic Bragg peaks above room temperature also in the undoped FeGa$_3$ indicating a complex magnetic structure \cite{gamza}. 

The $\mu$SR spectra show an evolution from short range order (in the $y=0.14$ and $y=0.17$ samples) to more long range order magnetism above $y\sim 0.20$. This is reflected in the field distribution and magnetic volume fraction probed by the polarized muons. The SRO is characterized by a broad field distribution centered around a zero internal field.  Moreover, magnetism develops only a partial volume fraction. By contrast above $y\sim 0.20$ the field distribution, while remaining broad, is characterized by the presence of a non-zero internal field  $B_{\mu}$, with the ZF spectra showing a heavily damped spontaneous spin precession. This indicates LRO of a magnetic ground state which, as shown by the weak TF measurements, develops in the full volume fraction. 

\begin{figure}[h!]
 \centering
 \includegraphics[width=9cm]{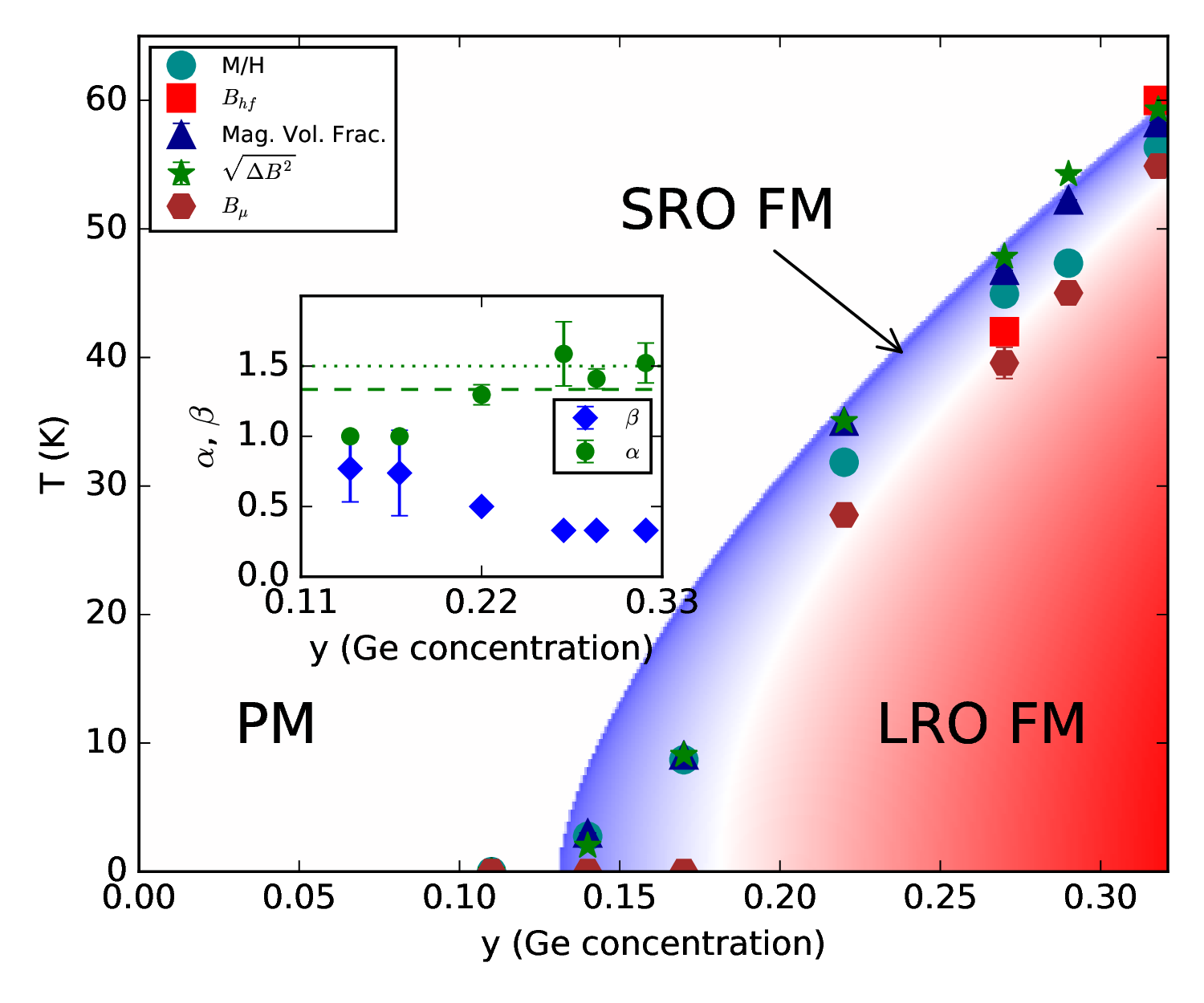}
 \caption{Phase diagram of FeGa$_{3-y}$Ge$_y$.  The transition temperatures as obtained from magnetization, $\mu$SR and MS analysis are shown.  The blue region corresponds to short range magnetic order, the red region correspond to long range magnetic order interactions.  In the inset the $\alpha$ and $\beta$ parameters obtained from the fits of the temperature dependence of the internal field and field width are shown.}\label{phasediag}
\end{figure}

The evolution of the character of the magnetic order with doping is also apparent in the temperature  dependence of the local magnetization expressed by the internal field measured by $\mu$SR and M\"ossbauer spectroscopy $B_{\mu}(T)$ and $B_{hf}(T)$, which were fitted with the generic expression 

\begin{equation}
B_{(\mu,hf)}(T)=B_0\left(1-\left(\frac{T}{T_C}\right)^{\alpha}\right)^{\beta}.\label{eq6}
\end{equation}
The curves with the best sets of parameters are plotted in Fig. \ref{fig2}(a)-(b) and Fig. \ref{figmoss} and the evolution of $\alpha$ and $\beta$ with $y$ is plotted in the insert of Fig. \ref{phasediag}. Because of the scarcity of points close to $T_{\rm C}$ we fixed or restrained in some cases $\beta$. It turns out that the choice of $\beta=\frac{1}{3}$, which reproduce the evolution of the magnetization close to $T_{\rm C}$ in most 3D magnetic systems approximately well\cite{nolting}, gives the best agreement with the experimental data in the region with LRO. This value is supported by the fact that $\alpha$ is then found to be very close to $\frac{3}{2}$ as expected in the case of low temperature contributions to spin fluctuations arising from spin wave excitations.  On lowering the Ge concentration there is a trend toward a reduction of $\beta$ and an increase of $\alpha$.  For $y=0.22$, the local field plotted in Fig. \ref{fig2}(a) is well described by $\beta=0.5$ and $\alpha=1.29(8)$, in agreement with the prediction from the SCR theory for spin fluctuations ($\beta=0.5$ and $\alpha=\frac{4}{3}$) \cite{moriya}.  

At lower doping where SRO is observed and $B_{\mu}=0$, we fitted $\lambda(T)$ and $\sigma(T)$, since both quantities are also a measure of the local magnetization. The temperature dependence is of the form $\left(1-\frac{T}{T_C}\right)^{\beta}$ with $\alpha=1$ and $\beta=0.74(30)$ for $y=0.17$ and $\beta=0.77(23)$ for $y=0.14$, which deviates from the SCR prediction. This doping region closer to the putative FMQCP displays other unusual behavior. The evolution of the magnetic fraction indicates an inhomogeneous  disappearance of the magnetic order in its vicinity, as indicated by the reduction of $V_m(0)$ near $y_c$.  The $\mu$SR results around $y=0.14-0.17$ evidence the coexistence of short-range ordered spin clusters coexisting with a non-magnetic environment. The combined Lorentz-Gauss field distribution indicate that the spin  structure of these clusters consists of a diluted distribution of larger magnetic moments embedded in a dense matrix of randomly distributed moments of smaller value.  This corresponds to the situation where a few Fe atoms possess a large moment in a sea of lower momenta Fe. The ratio $\frac{\sigma_T}{\lambda}\simeq\frac{1}{2}$, can be taken as a rough measure of the relative magnitudes of the two types of moments. 

This finding is in agreement with recent first-principle DFT calculations\cite{alvarez} which have indicated that the magnetic moments are not uniformly distributed throughout all Fe atoms as we would expect in a simple itinerant picture and that, depending on the doping concentration and the lattice distribution of the Ge dopants, different groupings of magnetic moments on the Fe atoms will form.  Qualitatively, the inhomogeneous situation around $y=\frac{1}{6}\simeq 0.17$ may be seen also as a consequence of the Fe coordination. Since each Fe has 6 Ga neighbors, around this concentration, on average,  one Ge atom will occupy a neighbor Ga site, inducing the small moment states and only in a few cases two Ge impurities close to Fe will induce a higher spin density. Specifically, in the DFT calculations different values of magnetic moments on the Fe atoms are predicted not only depending on their position with respect to the Ge impurity but also depending whether the impurity occupies a Ga1 or the more likely Ga2 position. For small dopant concentrations, some Fe sites are predicted to carry no moments. Our $\mu$SR measurements find a sizable non-magnetic volume fraction even at $y=0.17$. The local probe character of $\mu$SR puts a lower limit to the size of the non-magnetic regions to at least a few  lattice constants. This finding also supports the picture of a complex nature of the magnetism of FeGa$_{3-y}$Ge$_y$ with the exact lattice position and distribution of each Ge dopant directly influencing the appearance and site of the Fe spin.

The LRO order and the full magnetic fraction found on increasing the dopant distribution reflect the evolution to a more uniform magnetism with all Fe atoms having similar moments, which appears to be accompanied by a more itinerant character as indicated by nuclear quadrupolar resonance measurements \cite{majumder}. Interestingly, in most of the distributions leading to these fully ferromagnetic states, the induced magnetic moments are predicted to be oriented in the same direction \cite{alvarez}.  This is consistent with the present $\mu$SR and  MS results, suggesting a local hyperfine field lying perpendicular to the crystallographic $c$ axis.

Fig. \ref{phasediag} shows the phase diagram as obtained from the present $\mu$SR, MS and magnetization measurements. The temperatures defining the phase boundaries have been derived from the inflection point of the magnetization (Fig. \ref{magnetization}), from the 50 \% value of the magnetic volume fraction and from the onset temperature value where an internal field and broadening are detected by $\mu$SR and M\"ossbauer spectroscopy (Fig. \ref{fig2} and \ref{figmoss}).  No internal field $B_{\mu}$ has been detected at low doping. This together with the previously discussed field distribution is a strong indication of FM with SRO in this part of the phase diagram. SRO develops into LRO with increasing $y$. However, even at dopings where the low temperature state is long range ordered, magnetism appears at the thermal phase boundary first as SRO before gradually developing into the LRO state. This is also reflected by the onset temperature of the field broadening  being higher than the one of the internal field and is also reflected by a monotonic increase of $\frac{\sqrt{\Delta B^2}}{B_{\mu}}$ on approaching the Curie temperature from below (see Fig. \ref{fig2}(c)). 

\section{Conclusions}

In conclusion, magnetization, ZF, TF and LF $\mu$SR and M\"ossbauer spectroscopy measurements have been performed on FeGa$_{3−y}$Ge$_y$ singlecrystalline samples with $y$ ranging from 0.11 to 0.32. The $\mu$SR and MS spectra provide evidence for magnetism developing from short range order near the FMQCP to long range order with a heavily damped spontaneous precession showing up for $y=0.22$ and above. For the low dopant concentration $y=0.14$ and $0.17$ part of the sample remains in a non-magnetic state even at the lowest temperatures.  

The ZF data indicate that the magnetic moment formation, its size and the consequent character of the magnetic order, depends not only on the dopant concentration but also on details of the Ge dopant distribution as suggested by recent DFT calculations \cite{alvarez}.  The suppression of magnetism in a fraction of the sample volume for dopings close to the $y=0.13$ where a critical point is expected calls for more detailed study of the critical behavior by a local probe. The pressure induced decrease of the magnetic volume fraction for $y=0.14$ and of $T_{\rm C}$ in high doping samples \cite{umeo} suggest also that the critical behavior could be tuned by pressure. 

{\bf Acknowledgement:\/}  The research leading to these results has received funding from the European Community's Seventh Framework Programme (FP7/2007-2013) under grant agreement n.°290605 (PSI-FELLOW/COFUND), and from the Brazilian funding agencies CNPq, FAPESP (grant n. 2011/19924-2) and CAPES.  The authors thank H. Luetkens, J. C. Orain, A. Amato, R. Khasanov and Z. Shermadini for the support during the muon spin rotation experiments, and H. Micklitz for the discussions held.


\begin{thebibliography}{99}

\bibitem{dagotto}
Elbio Dagotto, Science {\bf 309}, 257-262 (2005).
\bibitem{hausermann}
Ulrich Hausermann, Magnus Bostr\"om, Per Viklund, \"Osten Rapp and Therese Bj\"orn\"angen, J. Solid State Chem.{\bf 165}, Issue 1, 94-99 (2002).
\bibitem{thermoelectric2}
Y. Hadano, S. Nazaru, M. A. Avila, T. Onimaru, T. Takabatake, J. Phys. Soc. Jpn. {\bf 78}, 013702 (2009).
\bibitem{thermoelectric6}
V. Ponnambalam, D. T. Morelli, J. Appl. Phys. {\bf 118\/}, 245101 (2015).
\bibitem{thermoelectric1}
C. S. Lue, W. J. Lai, Y. K. Kuo, J. Alloys Compd., {\bf 392}, 72-75 (2005).
\bibitem{amagai}
Y. Amagai, A. Yamamoto, T. Iida and Y. Takanashi, J. Appl. Phys {\bf 392}, 72-75 (2004).
\bibitem{thermoelectric3}
N. Haldolaarachchige, A. B. Karki, W. A. Phelan, Y. M. Xiong, R. Jin, J. Y. Chan, S. Stadler, D. P. Young, J. Appl. Phys. {\bf 109}, 103712 (2001).
\bibitem{thermoelectric4}
B. Ramachandran, K. Z. Syu, Y. K. Kuo, A. A. Gippius, A. V. Shevelkov, V. Y. Verchenko, C. S. Lue, Jour. Alloys Compd. {\bf 608}, 229-234 (2014).
\bibitem{thermoelectric5}
M. Wagner-Reetz, R. Cardoso-Gil, Yu. Grin, J. Electron. Mater. {\bf 43}, Issue 6, 1857-1864 (2014).
\bibitem{yin}
Z. P. Yin and W. E. Pickett, Phys. Rev. B {\bf 82}, 155202 (2013).
\bibitem{umeo}
K. Umeo, Y. Hadano, S. Narazu, T. Onimaru, M. A. Avila, T. Takabatake, Phys. Rev. B {\bf 86}, 144421 (2012).
\bibitem{haldolaarachchige}
N. Haldolaarachchige, J. Prestigiacomo, W. A. Phelan, Y. M. Xiong, G. McCandless, J. Y. Chan, J. F. DiTusa, I. Vekhter, S. Stadler, D. E. Sheehy, P. W. Adams, D. P. Young, arXiv:1304:1897.
\bibitem{singh}
D. J. Singh, Phys. Rev. B {\bf 88}, 064422 (2013).
\bibitem{michael}
M. Cabrera-Baez, E. T. Magnavita, R. A. Ribeiro, M. A. Avila, J. Electron. Mater., {\bf 43}, 1988-1992 (2014).
\bibitem{majumder}
M. Majumder, M. Wagner-Reetz, R. Cardoso-Gil, P. Gille, F. Steglich, Y. Grin, M. Baenitz, Phys. Rev. B {\bf 93}, 064410 (2016).
\bibitem{gamza}
M. B. Gamza, J. M. Tomczak, C. Brown, A. Puri, G. Kotliar, M. C. Aronson, Phys. Rev. B {\bf 89}, 195102 (2014).
\bibitem{verchenko}
V. Yu. Verchenko, M. S. Likhanov, M. A. Kirsanova, A. A. Gippius, A. V. Tkachev, N, E. Gervits, A. V. Galeeva, N. B\"uttgen, W. Kr\"atschmer, C. S. Lue, K. S. Okhotnikov, A. V. Shevelkov, J. Solid State Chem. {\bf 194}, 316-368 (2012).
\bibitem{gippius}
A. A. Gippius, V. Yu. Verchenko, A. V. Tkachev, N. E. Gervits, C. S. Lue, A. A. Tsirlin, N. B\"uttgen, W. Kr\"atschmer, M. Baenitz, M. Shatruk, A. V. Shevelkov, Phys. Rev. B {\bf 89} 104426 (2011).
\bibitem{alvarez}
J. C. Alvarez-Quiceno, M. Cabrera-Baez, R. A. Ribeiro, G. M. Dalpian, J. M. Osorio-Guill\'en, M. A. Avila, Phys. Rev. B {\bf 94}, 014432 (2016).
\bibitem{botana}
A. S. Botana, Y. Quan, W. E. Pickett, Phys. Rev. B {\bf 92} 155134 (2015).
\bibitem{likhanov}
M. S. Likhanov, V. Yu. Verchenko, M. A. Bykov, A. A. Tsirlin, A. A. Gippius, D. Berthebaud, A. Maignan, A. V. Shevelkov, J. Solid State Chem. {\bf 236} 166-172 (2016).
\bibitem{canfield}
P. C. Canfield and Z. Fisk, Phil. Mag. 65 1117 (1992).
\bibitem{ribeiro} 
R. A. Ribeiro, and M. A. Avila, Phil. Mag. 92 2492 (2012).
\bibitem{bittar}
E. M. Bittar, C. Capan, G. Seyfarth, P. G. Pagliuso, Z. Fisk, J. Phys.: Conf. Series {\bf 200}, 012014 (2010).
\bibitem{osorio}
J. M. Osorio-Guill\'en, Y. D. Larrauri-Pizarro, G. M. Dalpian, Phys. Rev. B {\bf 86}, 235202 (2012).
\bibitem{mondal}
D. Mondal, C. Kamal, S. Banik, A. Bhakar, A. Kak, G. Das, V. R. Reddy, A. Chakrabarti, T. Ganguli, arXiv:1606.04500.
\bibitem{moriya}
T. Moriya, Spin Fluctuations in Itinerant Electron Magnetism, Springer-Verlag Berlin Heidelberg GmbH (1985).  T. Moriya, J. Magn. Magn. Mater. {\bf 14}, 1-46 (1979).
\bibitem{zhang}
Y. Zhang, M. Imai, C. Michioka, Y. Hadano, M. A. Avila, T. Takabatake, and K. Yoshimura (unpublished).
\bibitem{adroja}
D. T. Adroja, A. D. Hillier, J. G. Park, W. Kockelmann, K. A. McEwen, B. D. Rainford, K-H Jang, C. Geibel, T. Takabatake, Phys. Rev. B {\bf 78}, 014412 (2008).
\bibitem{spehling}
J. Spehling, M. G\"unther, C. Krellner, N. Yeche, H. Luetkens, C. Baines, C. Geibel, H. H. Klauss, Phys. Rev. B {\bf 85} 140406(R) (2012).
\bibitem{amato}
A. Amato, Rev. Mod. Phys. {\bf 69} 1119 (1997).
\bibitem{musrbook}
A. Yaouanc, P. Dalmas de R\'eotier, Muon Spin Rotation, Relaxation, and Resonance: Applications to Condensed Matter, Oxford University Press (2011).
\bibitem{schroeder}
A. Schroeder, R. Wang, P. J. Baker, F. L. Pratt, S. J. Blundell, T. Lancaster, I. Franke, J. S. M\"oller, J. Phys.: Conf. Series {\bf 551}, 012003 (2014).
\bibitem{mosswin}
Z. Klencs\'ar, MossWin 4.0 Manual, version 2012.08.20.
\bibitem{litterst}
F. J. Litterst, private code.
\bibitem{nolting}
W. Nolting, A. Ramakanth, Quantum Theory of Magnetism, Springer-Verlag Berlin Heidelberg (2009).
\bibitem{fhamilt}
K. Ruebenbauer and T. Birchall, Hyperfine Interact. {\bf 7}, 125, (1979).
\bibitem{krishnamurthy}
V. V. Krishnamurthy, K. Nagamine, I. Watanabe, K. Nishiyama, S. Ohira, M. Ishikawa, D. H. Eom, T. Ishikawa, T. M. Briere, Phys. Rev. Lett. {\bf 88}, 046402 (2002).
\end{thebibliography}
\end{document}